\documentclass{PoS}

\usepackage{color}

\usepackage{graphicx}

\usepackage{verbatim}
\usepackage{amsmath}
\usepackage{amssymb}
\usepackage{subfigure}
\usepackage{url}

\usepackage{multirow}
\usepackage{threeparttable}
\usepackage{paralist}

\usepackage{color}
\definecolor{darkgreen}{rgb}{0,0.5,0}

\newcommand{\SU}{\text{SU}}

\DeclareRobustCommand{\Fig}[1]{Fig.~\ref{#1}}

\DeclareRobustCommand{\Eq}[1]{Eq.~(\ref{#1})}

\newcommand{\be}{\begin{equation}}
\newcommand{\ee}{\end{equation}}

\newcommand{\bea}{\begin{eqnarray}}
\newcommand{\eea}{\end{eqnarray}}

\title{Theory Advances in BSM Physics}

\ShortTitle{Theory Advances in BSM Physics}

\author{\speaker{Matthew McCullough}\thanks{I would like to thank Gian Giudice and Peter Graham for comments on this draft and the participants of the CERN-CKC TH Institute on Neutral Naturalness and of the MIAPP `Indirect Searches' and `Anticipating 14 TeV' programmes for a number of enlightening discussions  which influenced the presentation of material in this talk.}\\
        Theory Division, Physics Department, CERN, CH 1211 Geneva 23, Switzerland\\
        E-mail: \email{matthew.mccullough@cern.ch}}

\abstract{Rather than attempting to summarise the full spectrum of recent advances in Beyond the Standard Model (BSM) theory, which are many, in this talk I will instead take the opportunity to focus on two frameworks related to the hierarchy problem currently receiving significant attention.  They are the `Twin Higgs' and the `Relaxion'.  I will summarise the basic underlying structure of these theories at a non-expert level and highlight some interesting phenomenological signatures or outstanding problems.}

\FullConference{International Symposium on Lepton Photon Interactions at High Energies\\
		17-22 August 2015\\
		University of Ljubljana, Slovenia}
		
\begin{document}

\section{Introduction}
\label{sec:introduction}
In the wake of Run I of the LHC the core puzzles which force us to look beyond the Standard Model remain.  The hierarchy problem, which is the focus of this talk, has become ever more acute.  The central issue of the hierarchy problem is that the mass-squared parameter of the Standard Model, which determines the weak scale, is quadratically sensitive to new physics at higher energies.  For the Standard Model taken in isolation this would not pose any problem as this parameter could be taken as an input.  This might tempt one to suggest that there is only the Standard Model and thus side-step any potential issue.  However gravity alone, and in particular the scale where quantum effects in gravity become important (the Planck scale), suggests the existence of new physics at energy scales $M_P \sim 10^{18}$ GeV which would feed into the Higgs mass and the weak scale, contradicting the observed hierarchy $v = 246 \text{ GeV} \ll M_P$.  Other suggestions for new physics scales arise in the form of the Landau pole for Hypercharge, the Majorana mass required for the neutrino mass See-Saw mechanism, the symmetry breaking scale in Grand Unified Theories, the Peccei-Quinn breaking scale, the inflationary scale, and so on.  Thus the ``Standard Model Only'' option is seemingly unavailable.

Another logical possibility is that the new physics at higher energy scales, such as quantum gravity, does exist but does not for some reason destabilise the Higgs mass.  If this were the case then the Higgs mass may again be taken as an input parameter, or with more machinery may be generated dynamically in some way.  In essence this would correspond to a boundary condition on the theory such that the Higgs mass parameter is extremely small at e.g.\ the Planck scale and receives no large contributions.  Unfortunately, it has not been robustly demonstrated that there can exist such a theory of quantum gravity which finds a loophole in the simple Wilsonian arguments that lead to our picture of the hierarchy problem.  Thus this class of possibilities is currently unsatisfying and, although an interesting and worthwhile area of active investigation,\footnote{See e.g.\ \cite{Giudice:2014tma} and references contained within.} does not yet exist as a robust solution of the hierarchy problem.

It would seem then that the only option left on the table is to take the hierarchy problem at face value and to try and solve it at energies below the Planck scale within field theory.  A handful of options for solving the hierarchy problem have been proposed over the last few decades and among them a leading possibility is Supersymmetry.  Supersymmetry enforces a fermionic symmetry which ties the mass of the Higgs scalar to the mass of a fermion, the `Higgsino'.  Since fermions enjoy chiral symmetries which protect their mass, and the scalar mass is now tied to the fermion mass, one can see how this symmetry can solve the hierarchy problem.  However Supersymmetry now faces serious phenomenological issues.

\begin{figure}[t]
\centering
\includegraphics[height=2.5in]{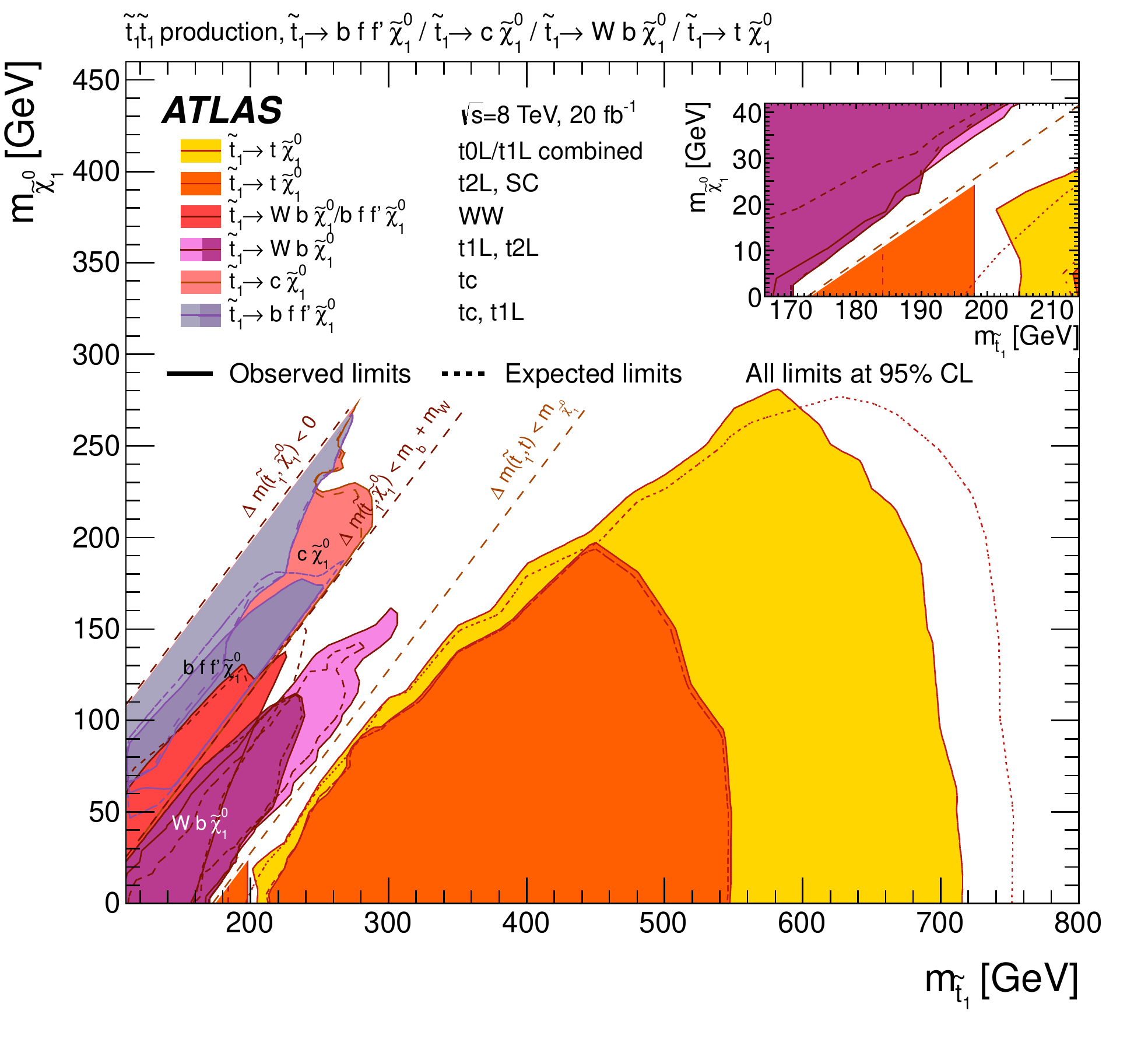} \qquad  \includegraphics[height=2.5in]{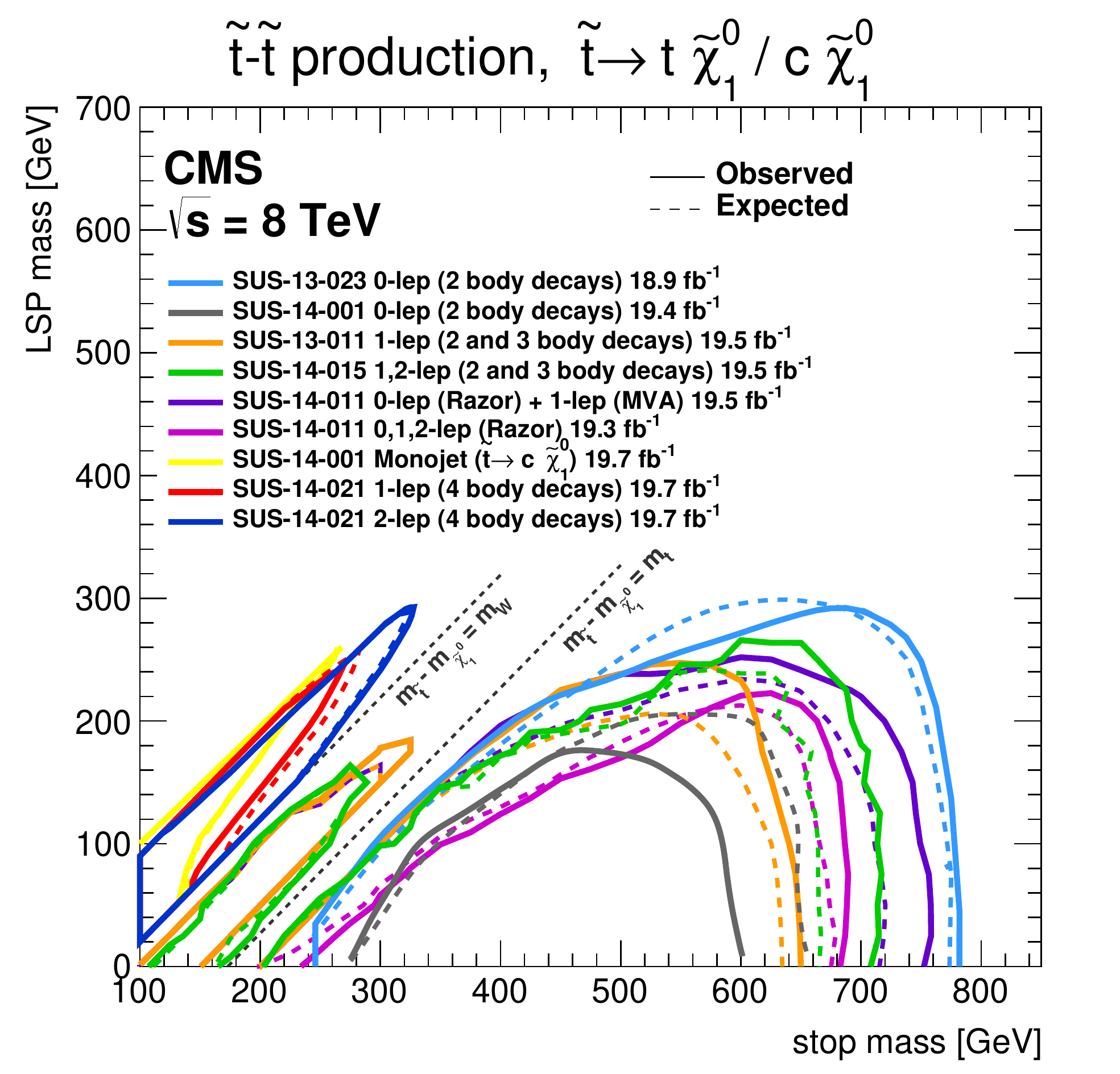}
\caption{Collated limits on direct stop squark production in a simplified model from ATLAS \cite{Aad:2015pfx} (left) and CMS \cite{CMS} (right).}
\label{fig:stops}
\end{figure}

During Run I of the LHC a vast array of measurements directly explored many corners of high energy Standard Model phenomenology and, with the exception of a few tentative hints for new physics which remain to be investigated during Run II, the Standard Model continues to reign supreme.  No evidence for a resolution of the hierarchy problem has yet arisen.  With regard to Supersymmetry the simplified limits on stop squarks shown in \Fig{fig:stops} paint a lucid picture.  To convincingly solve the hierarchy problem we would expect the stop squark masses to lie below scales of $\sim 400$ GeV or so.  However we see in \Fig{fig:stops} that much of this parameter space is already ruled out in a simplified model.  Of course it is possible that the stop squarks and other SUSY resonances are light but have evaded detection thus far due to a variety of different mechanisms.  However it is becoming ever more difficult to dodge the conclusion that a plethora of figures such as \Fig{fig:stops} from LHC Run I are telling us that the most basic ideas of Supersymmetric naturalness are unlikely to be realised in nature.

\begin{figure}[t]
\centering
\includegraphics[height=2.5in]{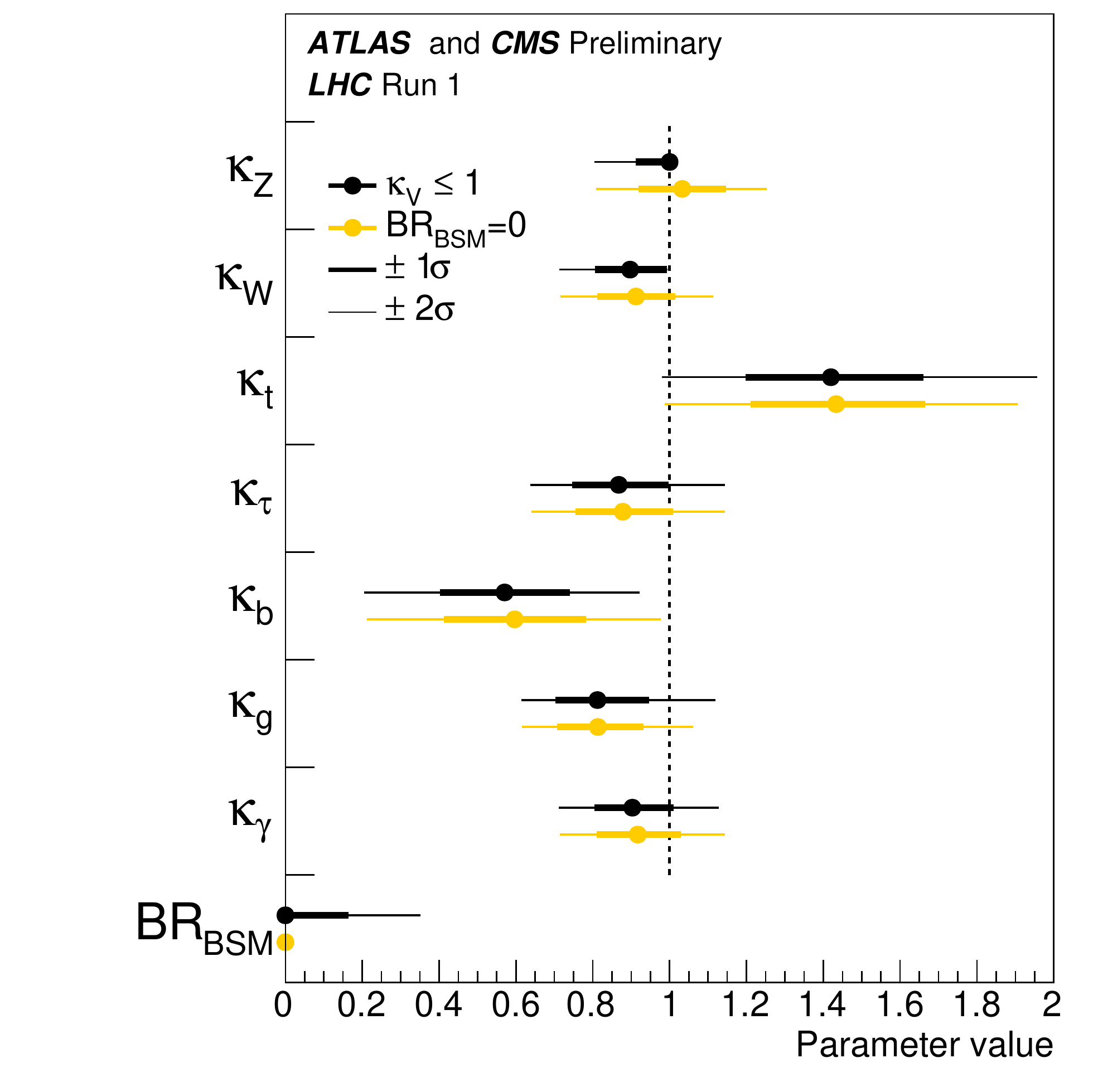}
\caption{Combined Higgs coupling constraints from ATLAS and CMS \cite{ATLAS-CONF-2015-044}.}
\label{fig:higgs}
\end{figure}

In addition to the direct searches for new physics, the discovery of the Higgs boson during Run I has also cast stark light on the nature of many solutions to the hierarchy problem.  In Supersymmetry the Higgs boson is expected to be reasonably light at tree level $m_h^2 \lesssim M_Z^2$, thus the measured mass $m_h = 125$ GeV points towards one loop corrections, mostly from stop squarks, that are comparable to the tree level contribution.  This pushes Supersymmetry into uncomfortable territory where heavy stops squarks are required and significant tuning is a consequence.  Furthermore, the measured couplings of the Higgs boson, shown in \Fig{fig:higgs}, already suggest a Higgs boson with properties similar to the Standard Model Higgs.  However in natural solutions to the hierarchy problem the Higgs boson couplings will typically be modified, often by an $\mathcal{O}(1)$ factor.  These observations indirectly put pressure on the possible structure of a solution to the hierarchy problem.  All-in-all, while conventional Supersymmetric models may still be realised in nature, this compelling theoretical picture is now in tension with the empirical picture that has emerged from Run I of the LHC.

What does all of this mean for the hierarchy problem?  First of all, it is clear that the lack of observation of BSM physics at Run I does not mean that the hierarchy problem has somehow vanished.  Rather it is now more acute, more formidable, a more pressing challenge.  In response, theorists are reluctant to surrender to potentially flimsy speculations about the possible behaviour of unknown physics at high energies.  On the contrary, the LHC Run I observations have stimulated a fresh wave of creativity on the theoretical front, of which I will now attempt to review a subset.  Due to limitations of time in this talk I will focus only on a radical old idea, the `Twin Higgs' \cite{Chacko:2005pe} mechanism which is receiving renewed attention, and on a radical new idea the `Relaxion' \cite{Graham:2015cka}, which may open the door to an entirely new class of solutions to the hierarchy problem intimately tied to the cosmology of the early Universe.

\section{The Twin Higgs}
At its heart the Twin Higgs \cite{Chacko:2005pe} is a Pseudo-Nambu-Goldstone (PNG) Higgs model.  I will attempt to present the model from this perspective.  Goldstone's theorem tells us that when a continuous global symmetry is spontaneously broken there will exist massless degrees of freedom.  This is immediately suggestive of a candidate structure for solving the hierarchy problem, since it may allow scalar particles to be significantly (and naturally) lighter than other mass scales in the theory.  Let us now consider a practical example that we will grow piecewise into the Twin Higgs

We begin with a scalar multiplet $H$ transforming as a fundamental under a global $\SU(4)$ symmetry.  The renormalizable potential for this theory is
\be
V = - m^2 |H|^2 + \frac{\lambda}{2} |H|^4
\label{eq:th}
\ee
where we have intentionally written a negative mass-squared.  In the vacuum the global symmetry breaking pattern is $\SU(4) \to \SU(3)$, thus irrespective of the magnitude of $m$ there will exist $7$ massless Goldstone bosons.  It is important to keep in mind that $m$ could be very large and in a theory with new physics scales, $m^2$ will contain all of the perturbative divergent contributions.  For example, if there are new states of mass $\Lambda$ we expect contributions $m^2 \sim \text{loop} \times \Lambda^2$.  This does not introduce additional quadratic divergences to the mass of the Goldstone bosons since these contributions are $\SU(4)$ symmetric and thus the Goldstone boson masses are still protected by Goldstone's theorem.

Let us break up $H$ into a representation of $\SU(2)_A \times \SU(2)_B \subset \SU(4)$ as
\be
H = \left( \begin{array}{c} H_A \\  H_B  \end{array} \right)   ~~.
\ee
We may rewrite \Eq{eq:th} as
\be
V = - m^2 \left( |H_A|^2 + |H_B|^2 \right) + \frac{\lambda}{2} \left( |H_A|^2 + |H_B|^2 \right)^2 ~~,
\label{eq:symm}
\ee
which is precisely the same as \Eq{eq:th}, but written in a different manner. 

We now augment the theory by gauging the two $\SU(2)_A \times \SU(2)_B$ subgroups.  If the vacuum expectation value for $H$ lies completely in the $H_B$ field then the three Goldstone bosons from $H_B$ will be eaten by the $\SU(2)_B$ gauge bosons, to become their massive longitudinal components.  The four Goldstone bosons from $H_A$ will remain uneaten because the off-diagonal gauge bosons of $\SU(4)$ which would have eaten these degrees of freedom were explicitly removed from the theory when we chose not to gauge the full $\SU(4)$ symmetry.  Thus we have four light scalars charged under the unbroken $\SU(2)_A$ gauge symmetry.  It is apparent that if $\SU(2)_A$ could be identified with the SM weak gauge group, and if $H_A$ could be identified with the SM Higgs doublet, then we have a candidate solution of the hierarchy problem.  However there are some further complications which must first be overcome.

The first point to note is that by coupling the scalars to gauge bosons we have introduced a new source of quadratic divergences.  Regularising the theory with a cutoff $\Lambda$ we generate terms such as
\be
V \sim \frac{g_A^2}{16 \pi^2} \Lambda_a^2 |H_A|^2 + \frac{g_B^2}{16 \pi^2} \Lambda_a^2 |H_B|^2  ~~,
\label{eq:gaugeloop}
\ee
where as yet there is no reason to believe the effective cutoff is the same for each field.  However, if we impose an exchange symmetry on the entire theory $A \leftrightarrow B$ then $g_A = g_B$, and assume the UV physics respects this exchange symmetry, such that $\Lambda_a=\Lambda_b$, then the contributions in \Eq{eq:gaugeloop} are equal.  Furthermore, because they are equal they respect the $\SU(4)$ symmetry, thus they do not actually introduce any new quadratically divergent contributions to the Goldstone boson masses.  Hence these dangerous contributions have been ameliorated by a combination of Goldstone's theorem and the fact that an exchange symmetry accidentally enforces an $\SU(4)$-invariant structure on the quadratic part of the action.  This is worth reemphasising: quadratic divergences have not been removed from the theory, but the sensitivity of the Goldstone boson masses to those divergences has been removed by Goldstone's theorem.

Unfortunately at the level of the quartic couplings the picture is not as clean.  The scalar quartic couplings will run logarithmically due to the gauge interactions.  This running must only respect the exchange symmetry and the $\SU(2)_A \times \SU(2)_B$ symmetry, but not necessarily  the full $\SU(4)$ symmetry.  In practice, even if we enforce an $\SU(4)$ symmetric scalar potential in \Eq{eq:symm} at a scale $\Lambda$, at the lower scale of symmetry breaking $m$ we expect additional contributions to the effective potential
\be
V_{BR} \sim \frac{g_A^4}{16 \pi^2} \log \left( \frac{m}{\Lambda_a} \right)  |H_A|^4 + \frac{g_B^4}{16 \pi^2}  \log \left( \frac{m}{\Lambda_b} \right) |H_B|^4 ~~.
\ee
Even when we impose the exchange symmetry, $g_A = g_B$, $\Lambda_a=\Lambda_b$, these terms explicitly break the $\SU(4)$ symmetry, thus they will in general lead to a non-zero mass-squared for the now pseudo-Goldstone bosons
\be
m_{PNG}^2 \propto  \frac{g_A^4}{16 \pi^2} m^2 \log \left( \frac{m}{\Lambda} \right) ~~.
\label{eq:BR} 
\ee
This tells us that in this theory we may only hope to have a loop factor in the hierarchy $m_h^2 \sim  \frac{g_A^4}{16 \pi^2} m^2$, and, as $m$ is quadratically sensitive to the cutoff, a loop factor in the hierarchy $m^2 \sim  \frac{g_A^2}{16 \pi^2} \Lambda^2$.  In the end of the day with this mechanism we expect the cutoff scale of the full theory to be a weak loop factor above the weak scale, demonstrating that the Twin Higgs can only be a solution to the little hierarchy problem and to solve the full hierarchy problem this theory must be UV-completed.  Proposed scenarios for UV completion include a holographic model \cite{Geller:2014kta}, composite 4D models \cite{Barbieri:2015lqa,Low:2015nqa}, and SUSY models \cite{Chang:2006ra,Falkowski:2006qq,Craig:2013fga}.

A final issue is that the theory presented above respects the exchange symmetry $A \leftrightarrow B$.  This implies that the vacuum will also respect this symmetry, with $v_A = v_B$.  Amongst other things, this predicts that the SM Higgs boson would be a perfect admixture of $H_A$ and $H_B$ and would couple to the SM gauge bosons with a suppression factor $\cos \theta_{AB} = 1/\sqrt{2}$.  Clearly this is at odds with observations.  To resolve this issue let us now introduce a small soft symmetry breaking term
\be
V_{SB} = -m_B^2 |H_B|^2 ~~.
\label{eq:break}
\ee
This term explicitly breaks the global symmetry and even the exchange symmetry.\footnote{Although we are explicitly writing a term which breaks this symmetry, recently it has been shown that this breaking can arise spontaneously, with favourable implications for the tuning required in the theory \cite{Beauchesne:2015lva}.}  It is important to note that since $m_B$ breaks the exchange symmetry it may be small in a technically natural manner.  One way of seeing this is that it may be the only parameter which breaks this symmetry, thus all explicit exchange breaking must enter proportional to this parameter, hence all explicit breaking mass parameters are $m_{GB}^2 \propto m_B^2 \ll \Lambda^2$.  Even though the Goldstone bosons have obtained mass from this operator, this mass is insensitive to the cutoff and can be naturally small: $m_{GB} \ll \Lambda$, where $\Lambda$ can be interpreted as the cutoff of the theory.  Importantly, this exchange symmetry breaking can align most of the vacuum expectation value into the $B$ sector, realising $v_A \ll v_B$.  This will suppress the Higgs mixing with the other SM neutral scalars and will also allow a hierarchical structure $v_A \ll v_B \ll \Lambda$, at the cost of a tuning comparable to $v_B^2/v_A^2$.

As far as the scalar fields and the gauge interactions are concerned, this is essentially all that is required of the Twin Higgs model.  Hypercharge may be trivially included in this picture.  The last step is to couple the SM Higgs to fermions.  If we add Yukawa couplings of $H_A$ to fermions, for example the up quarks, as
\be
\mathcal{L} \supset \lambda_A H_A Q_A U^c_A ~~,
\ee
then we see an immediate problem.  The top quark loops lead to $\SU(4)$-violating quadratic divergences
\be
m_A^2 \propto \frac{\lambda_t^2}{16 \pi^2} \Lambda^2
\ee
and the solution has been destroyed.  However, the resolution is immediately apparent.  We enforce the exchange symmetry $A \leftrightarrow B$ by introducing Twin quarks with identical couplings, such that the Yukawa couplings are now
\be
\mathcal{L} \supset \lambda_A H_A Q_A U^c_A +\lambda_B H_B Q_B U^c_B
\ee
and the quadratic divergences are once again $\SU(4)$-symmetric
\be
V \sim \frac{\lambda_A^2}{16 \pi^2} \Lambda_a^2 |H_A|^2 + \frac{\lambda_B^2}{16 \pi^2} \Lambda_b^2 |H_B|^2  ~~,
\label{eq:toploop}
\ee
since $\lambda_A = \lambda_B$ and $\Lambda_A = \Lambda_B$.  Thus the theory at the scale $\Lambda$ is approximately $\SU(4)$ symmetric and the SM Higgs boson is realised as a pseudo-Goldstone boson of spontaneous global symmetry breaking.  This renders the Higgs boson naturally lighter than the UV cutoff of the theory, $m_h \ll \Lambda$.

We can also see that if the Twin symmetry is imposed for all degrees of freedom, including gluons and leptons, then at any loop order the Higgs mass will still be free of quadratic sensitivity to the cutoff.  This is the essence of the Twin Higgs mechanism which, in the simplest incarnation, requires an entire copy of the SM which is completely neutral under the SM gauge group, but with its own identical gauge groups.  The only communication between the SM and the Twin Sector is through the Higgs boson, as I will describe soon.

The presentation of the Twin Higgs mechanism may appear somewhat backwards and a little laborious in comparison to other possible presentations.  This has been intentional, in the hope that it may anticipate a potential misconception for those not familiar with Twin Higgs model.  It is sometimes considered that it is seemingly ad hoc or arbitrary to add an entire copy of the SM for the Twin Higgs mechanism to work.  Hopefully this section has made it clear that there is nothing arbitrary about the introduction of the new fields.  The mechanism is not justified by adding an entire copy of the SM and then proving a diagram-by-diagram, and loop-by-loop, cancellation of quadratic divergences.  Rather, the new fields are introduced in order to realise an exchange symmetry $A \leftrightarrow B$.  The exchange symmetry ensures that at the quantum level the quadratic part of the scalar potential respects an accidental $\SU(4)$ symmetry, even with quadratic divergences included.  The observed Higgs boson mass is insensitive to this $\SU(4)$-symmetric quadratic divergence because it is a pseudo-Goldstone boson of spontaneous $\SU(4)$ breaking.\footnote{It is also possible to see a diagram-by-diagram cancellation of quadratic divergences rather than relying on the symmetry-based argument here, however this is less illuminating.}  It is noteworthy that this structure has recently been generalised to the `Orbifold Higgs' of which the Twin Higgs is an example \cite{Craig:2014aea,Craig:2014roa}.  This generalisation opens up many new possibilities for the underlying symmetry structure of this class of models.

On purely \ae sthetic grounds the Twin Higgs model could be compared to other solutions to the hierarchy problem, or even to Supersymmetry with a modest tuning embraced.  The outcome of this comparison would dependent on the individual \ae sthete.  However, at least with respect to the hierarchy problem, nature has shown little regard for the \ae sthetic predilections of model builders and thus it is important that all interesting ideas for resolving the hierarchy problem, however little, be investigated at colliders.

\subsection{Phenomenology}
Compared to weak scale Supersymmetry, where the copious production of new coloured particles at the LHC is a generic prediction, the collider signatures of the Twin Higgs are thin on the ground.  In both theories a key prediction is the existence of so-called `top partners' which regulate the quadratically divergent top quark loops contributing to the Higgs mass-squared.  These top-partners must be close to the weak scale and are thus a prime target for testing either theory.  In Supersymmetry these are the coloured stop squarks and collider limits are already closing in on natural parameter space.  In the Twin Higgs these are fermions charged under Twin QCD$_T$ but not under SM QCD.  They are in fact the first known example of a theory with the moniker ``Neutral Naturalness'', used to describe theories in which the top-partners are not charged under QCD.  This drastically suppresses top-partner production at the LHC since the only coupling to the SM is through the Higgs and any top-partner production must go through an off-shell Higgs boson.  The most promising approaches to test the Twin Higgs lie elsewhere.

One very robust prediction of the Twin Higgs scenario is a universal suppression of Higgs couplings to SM states.  The reason for this is that the Higgs bosons from both sectors, $h_A$ and $h_B$, have a mass mixing controlled by the hierarchy of vevs $v_A^2/v_B^2$.  As $h_B$ is a SM singlet this is equivalent to the usual ``Higgs Portal'' mixing scenario where all Higgs couplings are diluted by a factor $\cos \theta$.  This mixing may be constrained by searching for an overall reduction in Higgs signal strengths at the LHC and, since the ratio $v_A^2/v_B^2$ is a driving indicator of the tuning in the theory, Higgs observations directly probe the tuning of the Twin Higgs scenario.  In fact, one-loop LEP constraints on modified Higgs couplings already push this tuning to the $\sim 10 - 20 \%$ level \cite{Burdman:2014zta}.

Another possibility is that due to the Higgs Portal mixing the heavy Higgs boson may be singly produced at the LHC and it could decay in SM states with signatures, but not signal strengths, identical to a heavy SM Higgs boson.  It may also decay to pairs of Higgs bosons, leading to resonant di-Higgs production.

\begin{figure}[t]
\centering
\includegraphics[height=1.3in]{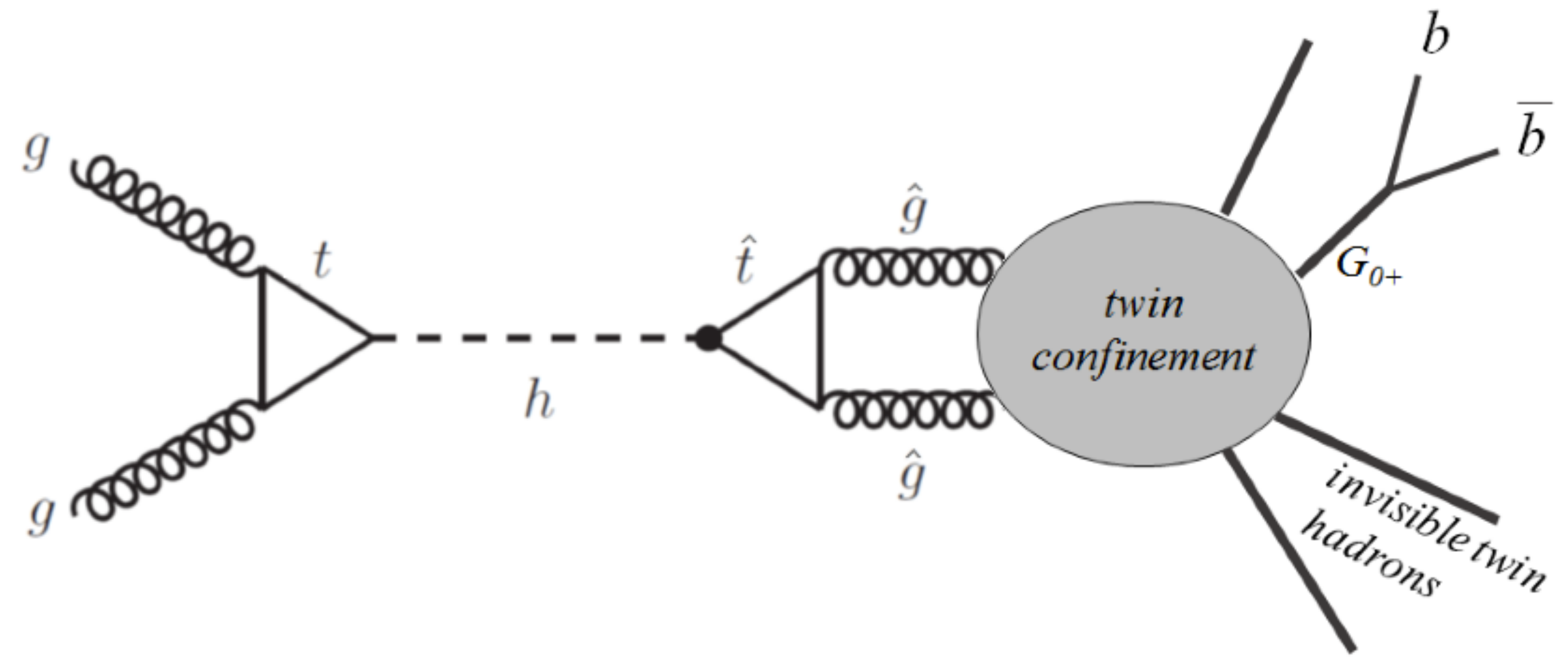} 
\caption{Twin glueball production and decay through the Higgs portal.  Figure taken from \cite{Craig:2015pha}.}
\label{fig:glueball}
\end{figure}

More exotic signatures arise once the Twin sector is considered in full.  If Twin sector states are produced through the Higgs Portal they may decay into lighter Twin sector states, eventually cascading down to the lightest states within the Twin sector.  These lightest states may then decay back into SM states, leading to a huge variety of exotic signatures.  In essence, the Twin Higgs scenario provides a framework in which many so-called `hidden valley' signatures \cite{Strassler:2006im,Strassler:2006ri,Han:2007ae} may be realised.  As the motivation comes from the hierarchy problem, it is necessary that the new states must lie within some proximity to the weak scale.  Taking naturalness as a guide there are many possibilities for the spectrum in the Twin sector since it is possible that the lighter states which are less relevant for Higgs naturalness may have modified couplings to the Twin Higgs or may even not exist in the theory, as in the `Fraternal Twin' scenario \cite{Craig:2015pha}.\footnote{With more radical violations of the underlying structure of the theory even more exotic possibilities arise, such as the Twin Top quarks acting as the right-handed neutrinos of the visible sector \cite{Batell:2015aha}.}

\begin{figure}[t]
\centering
\includegraphics[height=2.5in]{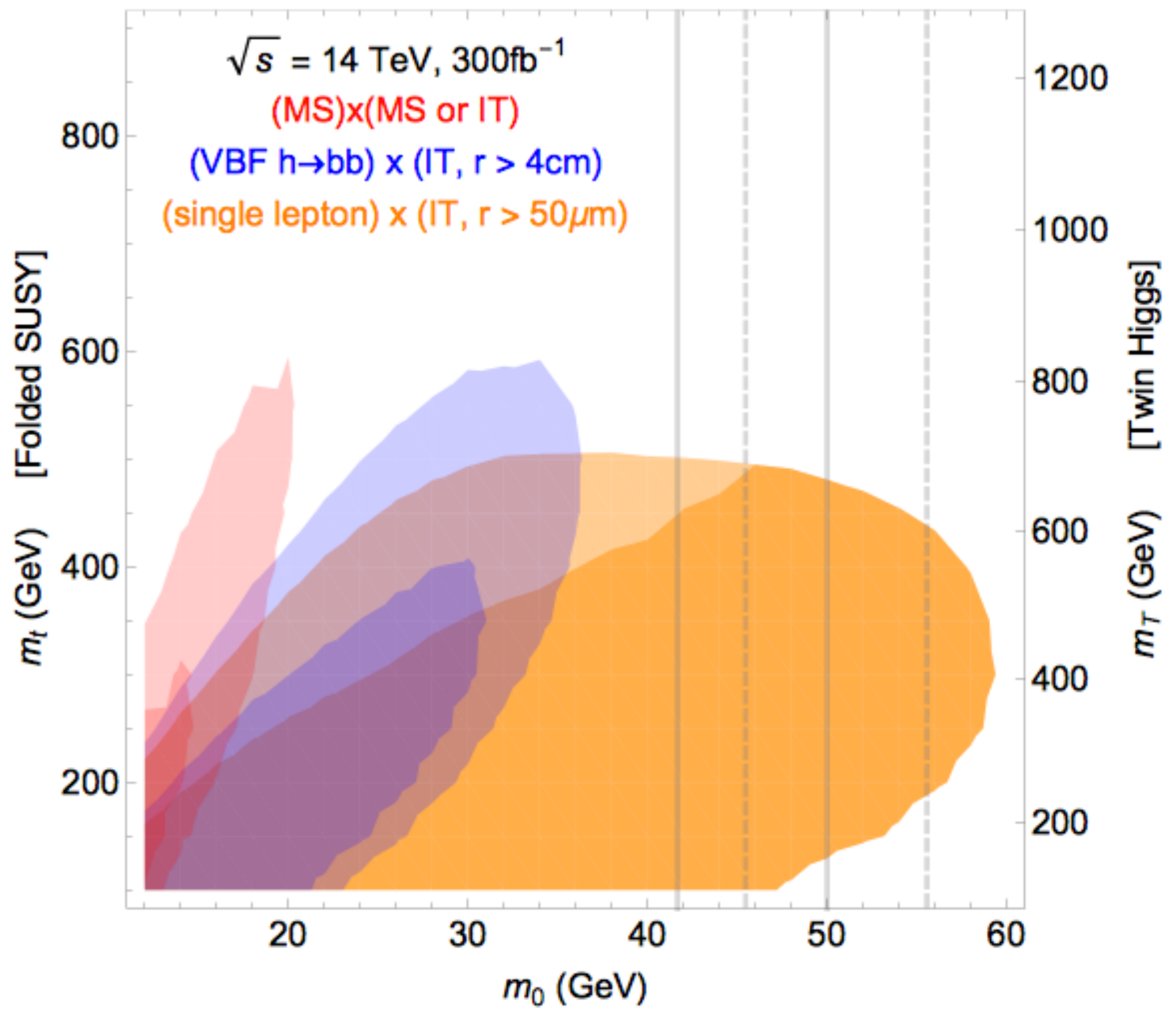}
\caption{Expected collider limits on the parameter space of the Twin Higgs model (far right axes) from constraints on exotic Higgs decays.  Figure taken from \cite{Curtin:2015fna}.}
\label{fig:limits}
\end{figure}

A particularly interesting example is for exotic Higgs decays into Twin glueballs, as depicted in \Fig{fig:glueball}.  This is possible because the Higgs couples to the Twin Top quarks, leading (at one loop) to a coupling to Twin gluons.  The Higgs may thus decay to the Twin glueballs, which then decay back through an off-shell Higgs to SM states, including bottom quarks.  Such an exotic Higgs decay signature can be used to search for the Twin sector states.  The expected reach for scenarios like this, taken from \cite{Curtin:2015fna}, is shown in \Fig{fig:limits}.

\subsection{Dark Matter}
Finally, it would be remiss to omit any discussion of dark matter in Twin Higgs models, since recent work has shown that the Twin Higgs setup is very attractive in this respect.  One observation is that a number of states in the Twin sector have weak-scale masses, such as the Twin $\tau_T$ lepton and also the Twin weak gauge bosons $Z_T,W^\pm_T$.  Furthermore, these states couple to other states in the Twin sector with Twin weak gauge couplings, which by the Twin symmetry are comparable to the SM weak gauge couplings.  It is not surprising then that if any of these states are cosmologically stable the Twin Higgs framework offers up excellent candidates to satisfy the so-called `WIMP Miracle'.  In fact, it has been shown that in variants of the Twin Higgs the Twin $\tau_T$ lepton and the Twin $W^\pm_T$ bosons, and even stable Twin glueballs, are valid dark matter candidates that may realise the observed dark matter density through thermal freeze-out while satisfying observational constraints \cite{Garcia:2015loa,Craig:2015xla}.

The picture for asymmetric dark matter is equally compelling.  It has long been known that the dark matter density may arise from an asymmetry in number density between dark baryons and dark anti-baryons, rather than originating from thermal freeze-out.  Furthermore, it may be that the dark number asymmetry is linked to the visible sector baryon number asymmetry if both are generated from the same mechanism in the early Universe.  Although we will not go into the details here, it is relatively straightforward to realise a scenario to predict a relationship between the visible baryon and dark baryon number asymmetries.  If these number asymmetries are the same then a key prediction is the mass of the dark matter, $m_{DM} \sim 5$ GeV.  This is an intriguing possibility.  However, unless a model robustly predicts this value for the mass, the observed dark matter density is not specifically predicted by the model, although it is accommodated.  However, in the Twin Higgs scenario a new confining gauge group, Twin $\text{QCD}_\text{T}$, is predicted and as the Twin sector states are heavier than the SM states it may run to strong coupling a little sooner than SM QCD.  In this case there is a strong motivation for stable Twin baryons to have mass in the ballpark of $5$ GeV and models realising this possibility have been found \cite{Garcia:2015toa,Farina:2015uea}.

\section{The Relaxion}
Recently a radical new approach to the hierarchy problem, ``The Relaxion'', has been proposed \cite{Graham:2015cka}.\footnote{A similar idea was considered much earlier for the cosmological constant problem \cite{Abbott:1984qf}, and alternative relaxation-based approaches to the gauge hierarchy problem have also been explored \cite{Dvali:2003br,Dvali:2004tma} more recently.}  As emphasised in \cite{Graham:2015cka}, if the Higgs is a fundamental scalar then the hierarchy problem relates to the fact that if we keep the theory fixed but change the Higgs mass, the point with a small Higgs mass is not a point of enhanced symmetry.  However, this may be a special point with regard to dynamics, since this is the point where the SM fields become light.

This perspective suggests that theories may exist where the Higgs mass is an evolving parameter in the early Universe.  Once the Higgs mass-squared becomes very small, or passes through zero, some non-trivial dynamics may occur which halts the evolution of the Higgs mass-squared, fixing it at a hierarchically small value.  This is precisely the form of the relaxion proposal.

The structure of the theory is relatively simple to write down.  Let us consider the SM as an effective theory at the scale $M$, which is the cutoff of the theory.  Following the standard EFT rules we include all of the operators, including non-renormalizable ones, consistent with symmetries.  All dimensionful scales are taken taken to the cutoff $M$.  We add to this theory a scalar $\phi$ which is invariant under a continuous shift symmetry, $\phi \to \phi + \kappa$, where $\kappa$ is some constant.  This shift symmetry only allows for kinetic terms for $\phi$.  We then add a dimensionful spurion field $g$ which breaks this shift symmetry.  As $g$ is the only source of shift symmetry breaking then a selection rule may be imposed, such that any potential terms for $\phi$ will enter in the combination $(g \phi / M^2)^n$.  Thus the theory is written
\be
\mathcal{L} = \mathcal{L}_{SM} - M^2 |H|^2 + g \phi |H|^2 + g M^2 \phi + g^2 \phi^2 + ...
\label{eq:relax}
\ee
where the ellipsis denote all of the other higher dimension terms and it should be understood that the coefficients of all the operators in \Eq{eq:relax} could vary by $\mathcal{O}(1)$ factors and the negative signs have been taken for ease of presentation.

The next step is to add an axion-like coupling of $\phi$ to the QCD gauge fields
\be
\frac{\phi}{32 \pi^2 f} G \widetilde{G} ~~.
\label{eq:ax}
\ee
This coupling is very special.  As $G \widetilde{G}$ is a total derivative, in perturbation theory \Eq{eq:ax} preserves the shift symmetry on $\phi$, thus it is consistent to include this operator without a factor of $g$ in the coupling.  Perturbatively this operator will not generate any potential for $\phi$, thus all of the shift-symmetry breaking terms involving $g$ remain radiatively stable and it is technically natural for them to be small.  However, non-perturbatively the full topological structure of the QCD vacuum breaks the shift symmetry $\phi \to \phi + \kappa$ down to a discrete shift symmetry $\phi \to \phi + 2 \pi f z$, where $z$ is an integer.  Thus the complete story is again one of symmetries.  $\phi$ enjoys a shift symmetry which is broken to a discrete shift symmetry by QCD effects.  The discrete shift symmetry is then broken completely by $g$.

The final trick lies in the fact that the $\phi$-potential generated by QCD effects depends on the light quark masses, which in turn depend on the Higgs vacuum expectation value.  In practice this potential will be
\bea
V_{QCD} & \sim & f_\pi^2 m_\pi^2 \cos \phi/f  \\
& \propto & f_\pi^3 m_q \cos \phi/f  \\
& \propto & f_\pi^3 \lambda_{u,d} \langle |H| \rangle \cos \phi/f ~~.
\label{eq:axion}
\eea
This dependence can be understood from the fact that if one quark $q$ were massless, then an anomalous chiral rotation $q \to e^{i \phi/f} q$ could eliminate $\phi$ from \Eq{eq:ax} and it would not reappear in a quark mass term.

Let us now consider the vacuum structure of the theory for two values of $\phi$.  If $M^2  - g \phi > 0$ then the effective Higgs mass-squared is positive.  QCD effects will break electroweak symmetry, and quark condensation will lead to a tadpole for the Higgs field, which will in turn lead to a very small vacuum expectation value for the Higgs.  Thus in this regime the axion potential of \Eq{eq:axion} exists but is extremely suppressed.  If $M^2  - g \phi < 0$ the effective Higgs mass-squared will be negative and the Higgs will obtain a vacuum expectation value.

\begin{figure}[t]
\centering
\includegraphics[height=2.0in]{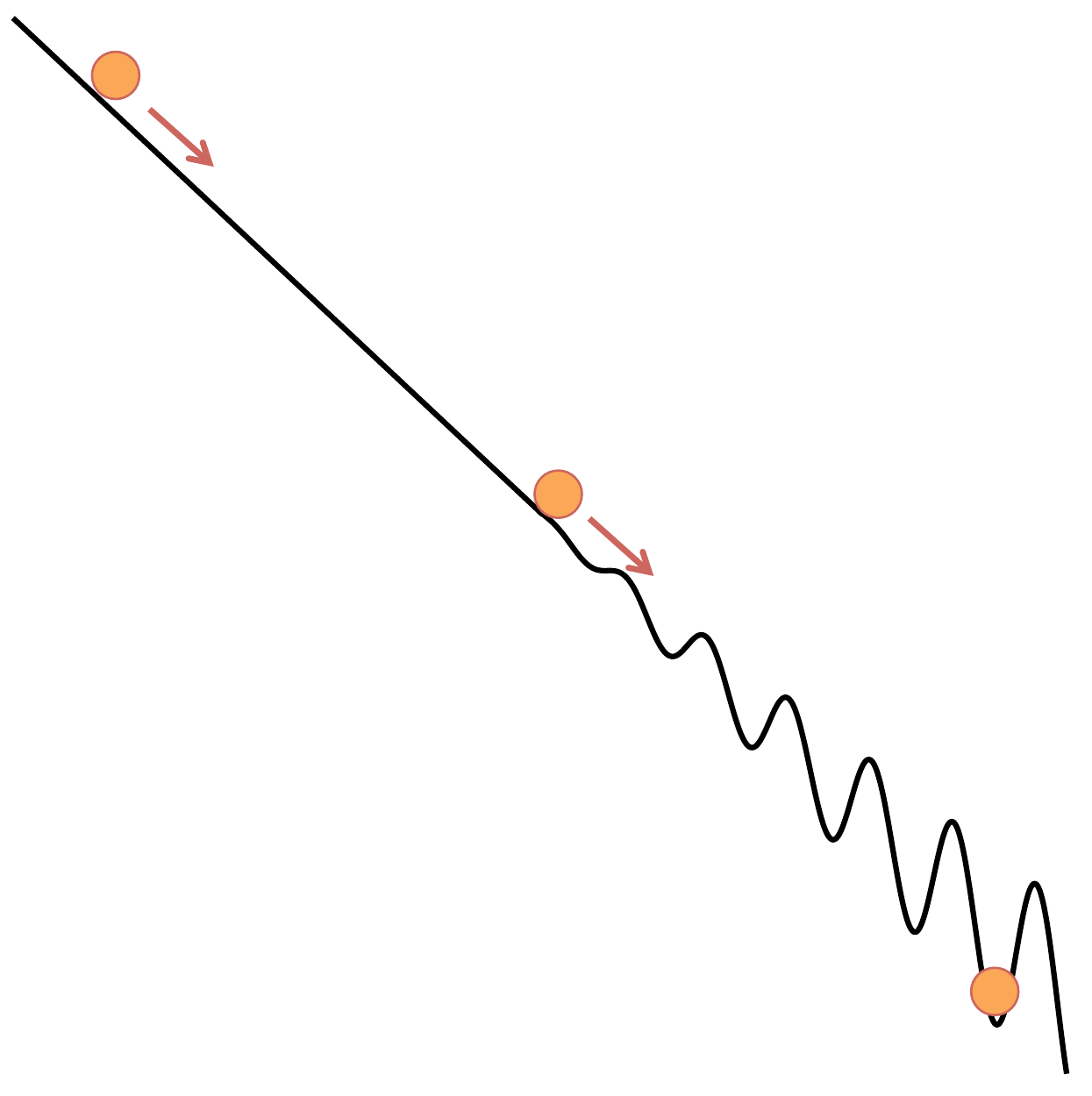}
\caption{Evolution of the relaxion field in the early Universe from a point where the effective Higgs mass-squared is postive (left), passing through zero (middle), and negative (right).}
\label{fig:relaxion}
\end{figure}

The general idea of the relaxion mechanism is sketched in \Fig{fig:relaxion}.  Imagine at the beginning of a period of inflation the relaxion field begins at values far from its minimum.  We can, without loss of generality, take this to be at $\phi = 0$.  Due to its potential it will roll, with Hubble friction providing the necessary dissipation for this to occur in a controlled manner.  All the while the effective Higgs mass-squared is evolving.  Once the effective mass-squared passes through zero the Higgs will obtain a vacuum expectation value and the axion potential of \Eq{eq:axion} will turn on, growing linearly with the Higgs vev.  If the gradient of this potential becomes locally great enough to overcome the gradient of the $g$-induced relaxion potential, i.e.\
\be
\frac{f_\pi^3}{f} \lambda_{u,d} \langle |H| \rangle \sin \phi/f > g M^2 ~~,
\ee
then the relaxion will stop rolling and become stuck.  Once it has become stuck the effective Higgs mass-squared has also stopped evolving.  If $g$ is taken to be appropriately small, then this evolution will cease at a point where the Higgs vev is small $\langle |H| \rangle \ll M$.  As $g$ is a parameter which can take values that are naturally small, and $g$ ends up determining the final Higgs vev, a naturally small value for the weak scale may be generated.\footnote{It should be noted that taking a parameter to be small is not a tuning if it is radiatively stable, i.e.\ if that parameter breaks a symmetry.  For example, the electron Yukawa coupling is a small number, but it is not a tuned number.}

If it could be taken at face value, the picture painted above is quite a beautiful portrait involving SM and BSM symmetries and dynamics.  QCD plays a crucial role in determining the weak scale and solving the hierarchy problem.  Only an axion-like field, already motivated by the strong-CP problem, is added.   Inflation, which is already required in cosmology, provides the dissipation required for solving the hierarchy problem.  We even find an explanation for some other puzzles in the SM, such as why there are some quark masses determined by the weak scale which are nonetheless lighter than the QCD strong coupling scale.  However, as we will see, some puzzles remain to be understood, presenting a number of interesting areas to explore on the theoretical front.

\subsection{Puzzles = Opportunities}
To determine the viability of the relaxion mechanism it is necessary to consider any constraints on the theory, which we will mostly take directly from \cite{Graham:2015cka} and list here.
\begin{itemize}
\item $\Delta \phi > M^2/g$:  For the relaxion to scan the entire $M^2$ of Higgs mass-squared it must traverse this distance in field space.
\item  $H_I > M^2/M_P$:  Inserting the previous $\Delta \phi$ into the potential we find that the vacuum energy must change by an amount $\Delta V \sim M^4$.  For the inflaton to dominate the vacuum energy during inflation we require $V_I > M^4$, which corresponds to the aforementioned constraint on the Hubble parameter during inflation.
\item  $H_I < \Lambda_{QCD}$:  For the non-perturbative QCD potential to form, the largest instantons, of size $l \sim 1/\Lambda_{QCD}$, must fit within the horizon.
\item $H_I < (g M^2)^{1/3}$:  Fluctuations in the relaxion field during inflation (due to finite Hubble) must not dominate over the classical evolution  if the theory is to predict a small weak scale.
\item $N_e \gtrsim H_I^2/g^2$:  Inflation must last long enough for the relaxion to roll over the required field range.
\item  $g M^2 f \sim \Lambda_{QCD}^4$:  It must be possible for a local minimum to form in the full relaxion potential whenever the Higgs vev is at the observed electroweak scale.
\end{itemize}
Combining these constraints it was found in \cite{Graham:2015cka} that the maximum allowed cutoff scale in the theory is
\be
M < \left( \frac{\Lambda^4 M^3_{Pl}}{f} \right)^{1/6} \sim 10^7 \text{ GeV} \times \left( \frac{10^9 \text{ GeV}}{f} \right)^{1/6} ~~.
\label{eq:cutoff}
\ee
It is compelling that such a large hierarchy can be realised within the relaxion framework.  On the other hand, since any new physics scales above this cutoff would introduce corrections to a fundamental scalar Higgs mass that are too large to be relaxed away within this framework, it is clear that a UV completion will be critical to understand the far ultraviolet story for the relaxion.  A Supersymmetric realisation was recently provided \cite{Batell:2015fma}.

Let us now saturate \Eq{eq:cutoff} and take $f=10^9$ GeV to explore the other parameters of the theory.  In this limit we find
\be
g \sim 10^{-26} \text{ GeV} ~~~,~ ~~  \Delta \phi \sim 10^{40} \text{ GeV} ~~~,~ ~~   5 \times 10^{-5} \text{ GeV} \lesssim H_I \lesssim 0.2 \text{ GeV}  ~~~, ~~~ N_e \gtrsim 10^{43} ~~.
\ee
All of these features are quite puzzling or unfamiliar.  As such they may represent interesting opportunities for continued theoretical investigation.  The parameter $g$ which explicitly breaks the shift symmetry is extremely small and may perhaps violate conjectures such as `gravity is the weakest force' \cite{ArkaniHamed:2006dz}.  Recent work along has already shed some light on this question \cite{Ibanez:2015fcv}.  On a related note, the required field displacement is not only large, it is `super-duper Planckian' \cite{Kaplan:2015fuy}  How such large field displacements can be accommodated by a story involving quantum gravity remains to be fully understood.  It is notable that the `Axion Monodromy' models for large field displacements arise within String Theory and can be used for the relaxion \cite{Ibanez:2015fcv}, so perhaps the mechanism can already be partly motivated and accommodated within theories of quantum gravity.

With regard to the inflationary aspects, the Hubble parameter is much smaller than is typical in inflationary models.  The number of e-foldings is huge.  Although not a problem in principle, it may be difficult to realise a natural inflationary model with the appropriate slow-roll parameters which reheats the Universe successfully and also accommodates the observed cosmological parameters, such as the spectral tilt.  These aspects have been explored recently within the context of explicit models \cite{Patil:2015oxa,DiChiara:2015euo}.  In addition, it has been shown that the mechanism can also be used at finite temperature, avoiding the necessity of an inflationary epoch \cite{Hardy:2015laa}.

A more tangible puzzle arises in the simplest QCD model presented above, as it is already excluded by experiment.  In the electroweak breaking vacuum the full relaxion potential will be minimized whenever
\be
\frac{\partial V_g}{\partial \phi} + \frac{\partial V_{QCD}}{\partial \phi} = 0  ~~,
\label{eq:min}
\ee
where $V_g$ is the scalar potential generated from the terms which explicitly break the shift symmetry, all originating from the parameter $g$, and $V_{QCD}$ is the axion-like potential coming from the non-perturbative QCD effects.  Since the relaxion is stopped by QCD effects before it reaches the minimum of $V_g$, the first term in \Eq{eq:min} is non zero.  This then implies that the second term in \Eq{eq:min} must also be non-zero.  By construction, $V_{QCD}$ is minimised whenever the effective strong-CP angle is zero, thus if it is not minimised the effective strong-CP angle must be non-zero.  In fact, it is typically expected to be close to maximal if the relaxion has stopped in one of the first minima that appears after the Higgs vev starts to grow.  This is in clear contradiction with experimental bounds on the strong-CP angle and so the model must be extended.

Options explored in \cite{Graham:2015cka} include an inflaton-dependent relaxion potential, such that the $g$-dependent terms effectively switch off after inflation terminates and the relaxion will roll to the nearest minimum of the axion potential, solving the strong-CP problem in the usual way.  Alternatively, the details of the QCD model described above may almost be mimicked by adding new fermions which are charged under a new strongly coupled gauge group.   If the dominant mass for the lightest fermion comes from the Higgs vev then a similar story plays out and it is not a problem that the effective $\theta$-angle for this new gauge group is predicted to be large.  Another model where the parameters of a similar construction also scan at the same time was presented in \cite{Espinosa:2015eda}, with a supersymmetric version in \cite{Evans:2016htp}.  Looking to the future, progress in rescuing the QCD model, or in developing new non-QCD models, may arise.

Finally, it was suggested that there may be a deeper structural issue with the relaxion if it is to be identified as the Goldstone boson of a spontaneously broken compact global symmetry.  The reason for this is that the $g$-dependent terms in the relaxion potential explicitly break a discrete gauge symmetry which would otherwise be preserved by the spontaneous symmetry breaking mechanism, essentially suggesting that in this case these potential terms cannot be generated within a local field theory \cite{Gupta:2015uea}.  However, there are a number of avenues that may alleviate this concern.  It may be that the relaxion is not a Goldstone boson.  At low energies it may appear as a scalar with an approximate shift symmetry with a UV completion that looks very different.\footnote{For recent work on the fate of shift symmetries in specific UV scenarios see \cite{Abel:2015rkm}.}  It may also be that, when the parameters of the potential are considered, the worrisome non-field-theory behaviour only enters at such high energies that you expect non-field-theory dynamics to be already present in any case \cite{Batell:2015fma}.  Furthermore, recently field theory realisations of relaxion monodromy have been put forward \cite{Ibanez:2015fcv,Hebecker:2015zss,Hebecker:2015tzo}, and it has also been shown that the relaxion potential may be realised within field theory where the effective periodicity with respect to the decay constant $f$ may be enhanced by a factor $e^N$, where $N$ is the number of axions in the model \cite{Choi:2015fiu,Kaplan:2015fuy}.  In summary, the somewhat exotic form of the relaxion potential may be a focal point for future work on the relaxion and there may be more to learn about large field displacements and quasi-periodic potentials.

What to make of the puzzles described here is a matter of perspective.  I would argue that the mechanism is original enough to warrant further investigation, and the various puzzles represent potential opportunities for theoretical progress.  We may stand to learn some very interesting things along the way.

\section{Summary}
By choosing to focus specifically on two topics of recent interest I have only scratched the surface of the wide array of recent advances in BSM physics.  However, if the health of a theoretical subfield is to be measured by its reaction to new experimental data and the generation of creative new theories, then I hope to have conveyed that BSM theory is in good shape.

\bibliographystyle{JHEP}
\bibliography{LPref}

\providecommand{\href}[2]{#2}\begingroup\raggedright\begin{thebibliography}{10}

\bibitem{Giudice:2014tma}
G.~F. Giudice, G.~Isidori, A.~Salvio, and A.~Strumia, {\it {Softened Gravity
  and the Extension of the Standard Model up to Infinite Energy}},  {\em JHEP}
  {\bf 02} (2015) 137, [\href{http://arxiv.org/abs/1412.2769}{{\tt
  arXiv:1412.2769}}].

\bibitem{Aad:2015pfx}
{\bf ATLAS} Collaboration, G.~Aad et~al., {\it {ATLAS Run 1 searches for direct
  pair production of third-generation squarks at the Large Hadron Collider}},
  {\em Eur. Phys. J.} {\bf C75} (2015), no.~10 510,
  [\href{http://arxiv.org/abs/1506.08616}{{\tt arXiv:1506.08616}}].

\bibitem{CMS}
{\it {CMS Supersymmetry Physics Results}},  tech. rep., CERN, Geneva, Sep,
  2015.

\bibitem{ATLAS-CONF-2015-044}
{\it {Measurements of the Higgs boson production and decay rates and
  constraints on its couplings from a combined ATLAS and CMS analysis of the
  LHC pp collision data at $\sqrt{s}$ = 7 and 8 TeV}},  Tech. Rep.
  ATLAS-CONF-2015-044, CERN, Geneva, Sep, 2015.

\bibitem{Chacko:2005pe}
Z.~Chacko, H.-S. Goh, and R.~Harnik, {\it {The Twin Higgs: Natural electroweak
  breaking from mirror symmetry}},  {\em Phys. Rev. Lett.} {\bf 96} (2006)
  231802, [\href{http://arxiv.org/abs/hep-ph/0506256}{{\tt hep-ph/0506256}}].

\bibitem{Graham:2015cka}
P.~W. Graham, D.~E. Kaplan, and S.~Rajendran, {\it {Cosmological Relaxation of
  the Electroweak Scale}},  \href{http://arxiv.org/abs/1504.07551}{{\tt
  arXiv:1504.07551}}.

\bibitem{Geller:2014kta}
M.~Geller and O.~Telem, {\it {Holographic Twin Higgs Model}},  {\em Phys. Rev.
  Lett.} {\bf 114} (2015) 191801, [\href{http://arxiv.org/abs/1411.2974}{{\tt
  arXiv:1411.2974}}].

\bibitem{Barbieri:2015lqa}
R.~Barbieri, D.~Greco, R.~Rattazzi, and A.~Wulzer, {\it {The Composite Twin
  Higgs scenario}},  {\em JHEP} {\bf 08} (2015) 161,
  [\href{http://arxiv.org/abs/1501.07803}{{\tt arXiv:1501.07803}}].

\bibitem{Low:2015nqa}
M.~Low, A.~Tesi, and L.-T. Wang, {\it {Twin Higgs mechanism and a composite
  Higgs boson}},  {\em Phys. Rev.} {\bf D91} (2015) 095012,
  [\href{http://arxiv.org/abs/1501.07890}{{\tt arXiv:1501.07890}}].

\bibitem{Chang:2006ra}
S.~Chang, L.~J. Hall, and N.~Weiner, {\it {A Supersymmetric twin Higgs}},  {\em
  Phys. Rev.} {\bf D75} (2007) 035009,
  [\href{http://arxiv.org/abs/hep-ph/0604076}{{\tt hep-ph/0604076}}].

\bibitem{Falkowski:2006qq}
A.~Falkowski, S.~Pokorski, and M.~Schmaltz, {\it {Twin SUSY}},  {\em Phys.
  Rev.} {\bf D74} (2006) 035003,
  [\href{http://arxiv.org/abs/hep-ph/0604066}{{\tt hep-ph/0604066}}].

\bibitem{Craig:2013fga}
N.~Craig and K.~Howe, {\it {Doubling down on naturalness with a supersymmetric
  twin Higgs}},  {\em JHEP} {\bf 03} (2014) 140,
  [\href{http://arxiv.org/abs/1312.1341}{{\tt arXiv:1312.1341}}].

\bibitem{Beauchesne:2015lva}
H.~Beauchesne, K.~Earl, and T.~Gregoire, {\it {The spontaneous $\mathbb{Z}_2$
  breaking Twin Higgs}},  \href{http://arxiv.org/abs/1510.06069}{{\tt
  arXiv:1510.06069}}.

\bibitem{Craig:2014aea}
N.~Craig, S.~Knapen, and P.~Longhi, {\it {Neutral Naturalness from Orbifold
  Higgs Models}},  {\em Phys. Rev. Lett.} {\bf 114} (2015), no.~6 061803,
  [\href{http://arxiv.org/abs/1410.6808}{{\tt arXiv:1410.6808}}].

\bibitem{Craig:2014roa}
N.~Craig, S.~Knapen, and P.~Longhi, {\it {The Orbifold Higgs}},  {\em JHEP}
  {\bf 03} (2015) 106, [\href{http://arxiv.org/abs/1411.7393}{{\tt
  arXiv:1411.7393}}].

\bibitem{Burdman:2014zta}
G.~Burdman, Z.~Chacko, R.~Harnik, L.~de~Lima, and C.~B. Verhaaren, {\it
  {Colorless Top Partners, a 125 GeV Higgs, and the Limits on Naturalness}},
  {\em Phys. Rev.} {\bf D91} (2015), no.~5 055007,
  [\href{http://arxiv.org/abs/1411.3310}{{\tt arXiv:1411.3310}}].

\bibitem{Craig:2015pha}
N.~Craig, A.~Katz, M.~Strassler, and R.~Sundrum, {\it {Naturalness in the Dark
  at the LHC}},  {\em JHEP} {\bf 07} (2015) 105,
  [\href{http://arxiv.org/abs/1501.05310}{{\tt arXiv:1501.05310}}].

\bibitem{Strassler:2006im}
M.~J. Strassler and K.~M. Zurek, {\it {Echoes of a hidden valley at hadron
  colliders}},  {\em Phys. Lett.} {\bf B651} (2007) 374--379,
  [\href{http://arxiv.org/abs/hep-ph/0604261}{{\tt hep-ph/0604261}}].

\bibitem{Strassler:2006ri}
M.~J. Strassler and K.~M. Zurek, {\it {Discovering the Higgs through
  highly-displaced vertices}},  {\em Phys. Lett.} {\bf B661} (2008) 263--267,
  [\href{http://arxiv.org/abs/hep-ph/0605193}{{\tt hep-ph/0605193}}].

\bibitem{Han:2007ae}
T.~Han, Z.~Si, K.~M. Zurek, and M.~J. Strassler, {\it {Phenomenology of hidden
  valleys at hadron colliders}},  {\em JHEP} {\bf 07} (2008) 008,
  [\href{http://arxiv.org/abs/0712.2041}{{\tt arXiv:0712.2041}}].

\bibitem{Batell:2015aha}
B.~Batell and M.~McCullough, {\it {Neutrino Masses from Neutral Top Partners}},
   {\em Phys. Rev.} {\bf D92} (2015), no.~7 073018,
  [\href{http://arxiv.org/abs/1504.04016}{{\tt arXiv:1504.04016}}].

\bibitem{Curtin:2015fna}
D.~Curtin and C.~B. Verhaaren, {\it {Discovering Uncolored Naturalness in
  Exotic Higgs Decays}},  \href{http://arxiv.org/abs/1506.06141}{{\tt
  arXiv:1506.06141}}.

\bibitem{Garcia:2015loa}
I.~Garcia~Garcia, R.~Lasenby, and J.~March-Russell, {\it {Twin Higgs WIMP Dark
  Matter}},  {\em Phys. Rev.} {\bf D92} (2015), no.~5 055034,
  [\href{http://arxiv.org/abs/1505.07109}{{\tt arXiv:1505.07109}}].

\bibitem{Craig:2015xla}
N.~Craig and A.~Katz, {\it {The Fraternal WIMP Miracle}},  {\em JCAP} {\bf
  1510} (2015), no.~10 054, [\href{http://arxiv.org/abs/1505.07113}{{\tt
  arXiv:1505.07113}}].

\bibitem{Garcia:2015toa}
I.~Garcia~Garcia, R.~Lasenby, and J.~March-Russell, {\it {Twin Higgs Asymmetric
  Dark Matter}},  {\em Phys. Rev. Lett.} {\bf 115} (2015), no.~12 121801,
  [\href{http://arxiv.org/abs/1505.07410}{{\tt arXiv:1505.07410}}].

\bibitem{Farina:2015uea}
M.~Farina, {\it {Asymmetric Twin Dark Matter}},
  \href{http://arxiv.org/abs/1506.03520}{{\tt arXiv:1506.03520}}.

\bibitem{Abbott:1984qf}
L.~F. Abbott, {\it {A Mechanism for Reducing the Value of the Cosmological
  Constant}},  {\em Phys. Lett.} {\bf B150} (1985) 427--430.

\bibitem{Dvali:2003br}
G.~Dvali and A.~Vilenkin, {\it {Cosmic attractors and gauge hierarchy}},  {\em
  Phys. Rev.} {\bf D70} (2004) 063501,
  [\href{http://arxiv.org/abs/hep-th/0304043}{{\tt hep-th/0304043}}].

\bibitem{Dvali:2004tma}
G.~Dvali, {\it {Large hierarchies from attractor vacua}},  {\em Phys. Rev.}
  {\bf D74} (2006) 025018, [\href{http://arxiv.org/abs/hep-th/0410286}{{\tt
  hep-th/0410286}}].

\bibitem{Batell:2015fma}
B.~Batell, G.~F. Giudice, and M.~McCullough, {\it {Natural Heavy
  Supersymmetry}},  \href{http://arxiv.org/abs/1509.00834}{{\tt
  arXiv:1509.00834}}.

\bibitem{ArkaniHamed:2006dz}
N.~Arkani-Hamed, L.~Motl, A.~Nicolis, and C.~Vafa, {\it {The String landscape,
  black holes and gravity as the weakest force}},  {\em JHEP} {\bf 06} (2007)
  060, [\href{http://arxiv.org/abs/hep-th/0601001}{{\tt hep-th/0601001}}].

\bibitem{Ibanez:2015fcv}
L.~E. Ibanez, M.~Montero, A.~Uranga, and I.~Valenzuela, {\it {Relaxion
  Monodromy and the Weak Gravity Conjecture}},
  \href{http://arxiv.org/abs/1512.00025}{{\tt arXiv:1512.00025}}.

\bibitem{Kaplan:2015fuy}
D.~E. Kaplan and R.~Rattazzi, {\it {A Clockwork Axion}},
  \href{http://arxiv.org/abs/1511.01827}{{\tt arXiv:1511.01827}}.

\bibitem{Patil:2015oxa}
S.~P. Patil and P.~Schwaller, {\it {Relaxing the Electroweak Scale: the Role of
  Broken dS Symmetry}},  \href{http://arxiv.org/abs/1507.08649}{{\tt
  arXiv:1507.08649}}.

\bibitem{DiChiara:2015euo}
S.~Di~Chiara, K.~Kannike, L.~Marzola, A.~Racioppi, M.~Raidal, and C.~Spethmann,
  {\it {Relaxion Cosmology and the Price of Fine-Tuning}},
  \href{http://arxiv.org/abs/1511.02858}{{\tt arXiv:1511.02858}}.

\bibitem{Hardy:2015laa}
E.~Hardy, {\it {Electroweak relaxation from finite temperature}},  {\em JHEP}
  {\bf 11} (2015) 077, [\href{http://arxiv.org/abs/1507.07525}{{\tt
  arXiv:1507.07525}}].

\bibitem{Espinosa:2015eda}
J.~R. Espinosa, C.~Grojean, G.~Panico, A.~Pomarol, O.~Pujolas, and G.~Servant,
  {\it {Cosmological Higgs-Axion Interplay for a Naturally Small Electroweak
  Scale}},  \href{http://arxiv.org/abs/1506.09217}{{\tt arXiv:1506.09217}}.

\bibitem{Evans:2016htp}
J.~L. Evans, T.~Gherghetta, N.~Nagata, and Z.~Thomas, {\it {Naturalizing
  Supersymmetry with a Two-Field Relaxion Mechanism}},
  \href{http://arxiv.org/abs/1602.04812}{{\tt arXiv:1602.04812}}.

\bibitem{Gupta:2015uea}
R.~S. Gupta, Z.~Komargodski, G.~Perez, and L.~Ubaldi, {\it {Is the Relaxion an
  Axion?}},  \href{http://arxiv.org/abs/1509.00047}{{\tt arXiv:1509.00047}}.

\bibitem{Abel:2015rkm}
S.~Abel and R.~J. Stewart, {\it {Shift-Symmetries at Higher Order}},
  \href{http://arxiv.org/abs/1511.02880}{{\tt arXiv:1511.02880}}.

\bibitem{Hebecker:2015zss}
A.~Hebecker, F.~Rompineve, and A.~Westphal, {\it {Axion Monodromy and the Weak
  Gravity Conjecture}},  \href{http://arxiv.org/abs/1512.03768}{{\tt
  arXiv:1512.03768}}.

\bibitem{Hebecker:2015tzo}
A.~Hebecker, J.~Moritz, A.~Westphal, and L.~T. Witkowski, {\it {Axion Monodromy
  Inflation with Warped KK-Modes}},  {\em Phys. Lett.} {\bf B754} (2016)
  328--334, [\href{http://arxiv.org/abs/1512.04463}{{\tt arXiv:1512.04463}}].

\bibitem{Choi:2015fiu}
K.~Choi and S.~H. Im, {\it {Realizing the relaxion from multiple axions and its
  UV completion with high scale supersymmetry}},
  \href{http://arxiv.org/abs/1511.00132}{{\tt arXiv:1511.00132}}.

\end{thebibliography}\endgroup

\end{document}